# On the Capability of Measurement-Based Quantum Feedback

Bo QI,   Hao PAN,   Lei GUO

Key Laboratory of Systems and Control, ISS, Academy of Mathematics and Systems Science,

Chinese Academy of Sciences, Beijing 100190, P. R. China

**Abstract**

As a key method in dealing with uncertainties, feedback has been understood fairly well in classical control theory. But for quantum control systems, the capability of measurement-based feedback control (MFC) has not been investigated systematically. In contrast to the control of classical systems where the measurement effect is negligible, the quantum measurement will cause a quantum state to collapse, which will inevitably introduce additional uncertainties besides the system initial uncertainty. Therefore, there is a complicated tradeoff between the uncertainty introduced and the information gained by the measurement, and thus a theoretical investigation of the capability of MFC is of fundamental importance. In this paper, inspired by both the Heisenberg uncertainty principle for quantum systems and the investigation of the feedback capability for classical systems, we try to answer the following three basic questions: (i) How to choose the measurement channel appropriately? (ii) Is the MFC still superior to the open loop control in dealing with the system information uncertainties? and (iii) What is the maximum capability or limitation of the MFC? These questions will be answered theoretically by establishing several theorems concerning the asymptotic reachability of eigenstates of a typical class of Hamiltonian control mode.

**Index terms** Quantum feedback control, quantum measurement, Heisenberg uncertainty principle, impossibility theorem.

I. INTRODUCTION

Over the past decades, much progress has been made in regulating the microscopic world, such as molecules, atoms and so on (e.g., [1], [2], [3]). The rapid development of the technology further promotes the development of the quantum control theory. According to whether or not the state information is used in regulating the quantum system during the control process, we may divide the control strategy into open-loop control (OLC) and feedback control (FC). For the OLC, we can further divide it into Hamiltonian and non-Hamiltonian control modes according to how the control law is introduced. For the Hamiltonian control mode (e.g., [4], [5], [6]), we mean the control is an adjustable parameter in the Hamiltonian. While in the non-Hamiltonian control case (e.g., [7], [8], [9]), one regulates the quantum system by adjusting some parameters of an auxiliary system, for example, the parameter of the environment of the system. For the FC, there are three types: (i) learning control (e.g., [10], [11], [12]) (ii) measurement-based feedback control (MFC) (e.g., [13], [14], [15], [16], [17]) and (iii) coherent feedback control (e.g., [18], [21], [22]). For the learning control, one begins with some initial control pulses, and then makes some measurement on the samples after the control process. A selected learning algorithm is used to adjust the control pulses and then the above steps are repeated on some new samples until the performance index does not change significantly. Hence, for this control mode, one needs different samples for different cycles. For the MFC mode, one performs some direct or indirect measurement on the quantum system and then design a control law based on the measurement information to regulate the quantum system. In contrast to the MFC mode, the coherent feedback control which was first put forward in [18] does not involve any measurement, and instead, the controller is coherently connected with the system plant and can be a quantum system itself. This type of feedback has inspired much subsequent research, see, (e.g., [19], [20]). Note that what we are really interested in is the state of the system plant. Therefore the controller itself will generally cause the quantum decoherence to the controlled system even though it coherently entangles with the system plant [21]. Thus whether the effect of the coherent feedback is better than the OLC for quantum systems needs to be investigated in depth.

In this paper, we only focus on the characteristics of the MFC of quantum systems, where measurement is needed to gain information of the system to be used for feedback control design. For classical systems, the back action effect of the measurement on the system state can be neglected, which, however, cannot be neglected in the quantum case. This is because, by the postulates of the quantum mechanics, the measurement on a quantum system will generally cause the quantum state collapse, which will inevitably introduce uncertainty to the system besides the original uncertainty that the system already shared [23]. Therefore, in the case of the quantum MFC, there are some fundamental questions which are not met in the classical control theory, for example,

- How to select the measurement channel appropriately? For example, how to gain the information of the system state as much as possible while the corresponding induced uncertainty to the system is as little as possible?
- Even if we have selected a measurement channel appropriately, the measurement induced uncertainty is still there, so whether the MFC is still superior to the OLC in dealing with the original uncertainty of the system is far from obvious and needs to be investigated seriously ([24], [25]).
- Inspired by the well-known Heisenberg uncertainty principle for quantum systems and the investigation of the maximum capability of feedback in classical systems (e.g., [26], [27], [28]), one may ask what is the maximum capability of the MFC for quantum systems?

In this paper we will give some partial, yet first rigorous, answers to these three fundamental questions respectively.

The paper is organized as follows. In Section 2, we briefly introduce the typical coherent control model and set up the control problems. In Section 3, we give a selection theorem on the measurement channel. Then we compare the capability of the OLC and the MFC in dealing with the initial state uncertainty in Section 4. We further give an impossibility theorem on the control capability of the MFC for quantum systems in Section 5. Section 6 concludes the paper with some remarks.

## II. THE COHERENT CONTROL MODEL

We now sketch the control model to be used (see e.g., ([16], [29]) for more details). Let $\mathcal{H}$ be the Hilbert space of the quantum system, which is of finite dimension $\dim \mathcal{H} = N < \infty$. The corresponding system state space will be represented by the convex set of density matrices

$$\mathcal{S} = \{\rho \in \mathbb{C}^{N \times N} : \rho = \rho^* \geq 0, \ \text{Tr}(\rho) = 1\},$$

where $\rho^*$ represents Hermitian conjugation of $\rho$, $\mathbb{C}^{N\times N}$ is the complex space of $N \times N$ and $\text{Tr}(\cdot)$ is the trace operator. The set of pure states in $\mathcal{S}$ can be represented as

$$\mathcal{S}_0 = \{\rho \in \mathbb{C}^{N\times N} : \rho = \psi\psi^*, \psi \in \mathbb{C}^N, \ \psi^*\psi = 1\}.$$

The quantum bath is modeled by the symmetric Fock space $\mathcal{F}$. Now, let $\mathcal{A}$ denote the von Neumann algebra generated by the set of all bounded operators on $\mathcal{H} \otimes \mathcal{F}$. Then the quantum probability space of the system and the bath is $(\mathcal{A}, \rho)$, where the state $\rho = \rho_{\mathcal{H}} \otimes \rho_{vacuum}$ is given by some state $\rho_{\mathcal{H}}$ on $\mathcal{H}$ and the vacuum state $\rho_{vacuum}$ on the bath $\mathcal{F}$.

Under some proper approximations, the quantum dynamics can be described by the Hudson-Parthasarathy equation[1], (e.g., [29], [30])

$$dU_t = \{-i(H_0 + u_t H_b)dt - \frac{1}{2}L^*Ldt + \sqrt{\kappa}LdA_t^* - \sqrt{\kappa}L^*dA_t\}U_t, \tag{1}$$
$$U_0 = I.$$

Here, $U_t$ describes the unitary evolution of the whole plant, $H_0$, $H_b$ and $L$ are system operators, and $A_t$ is the annihilation operator on $\mathcal{F}$. The symbol $*$ denotes the Hilbert space adjoint as well as the scalar complex conjugate. Furthermore, $H_0$ and $H_b$ are Hermitian and can be considered as the effective Hamiltonian and the control channel of the system respectively. $\{u_t\}$ is an adjustable parameter process and $\kappa > 0$ is the effective interaction strength.

In this paper, for the MFC, we will measure the field observable $Y_t = U_t^*(A_t + A_t^*)U_t$. Noting the form of the observable $Y_t$ and the quantum dynamics (1), $L$ may be called the measurement channel.

For any system observable $X$, let $j_t(X) = U_t^*(X \otimes \text{Id}_{\mathcal{F}})U_t$. Then from the quantum Itô rule and (1) one has

$$dY_t = \sqrt{\kappa}j_t(L + L^*)dt + dA_t^* + dA_t. \tag{2}$$

Now, we want to slightly extend the observation model to contain the case where the detection efficiency is not perfect. In fact, there are always some technical noises added to the measured signal in practice. We will model these noises by an additional term $B_t + B_t^*$ which is uncorrelated with $A_t + A_t^*$ and does not interact with the system. Therefore, one has

$$dY_t = \sqrt{\kappa\eta}j_t(L+L^*)dt + \sqrt{\eta}(dA_t^* + dA_t) + \sqrt{1-\eta}(dB_t + dB_t^*),$$

---

[1] We always work in units such that $\hbar = 1$.

where $\eta \in (0,1]$ is the detection efficiency, with $\eta = 1$ corresponds to a perfect detection.

We remark that $Y_t$ has the following two crucial properties [16]:

1) $Y_t$ is self-nondemolition, i.e., $[Y_t, Y_s] = 0$, for all $s < t$. This property implies that the observation process can be viewed as a stochastic process on a classical probability space by the spectral theorem [31].
2) $Y_t$ is nondemolition, i.e., $[j_t(X), Y_s] = 0$, for all $s < t$. This property implies that we can estimate the state of the system based on the observations.

Note that what we are really interested in is the state of the system. Thanks to the above properties of the observation process $Y_t$, we can use the quantum filtering theory to estimate the state of the system.

First, we calculate the quantum conditional expectation $\pi_t(X) = \mathcal{E}[j_t(X)|\mathcal{B}_t]$ where $\mathcal{E}$ is the quantum expectation, $\mathcal{B}_t$ is the $\sigma$-algebra generated by $Y_{s \leq t}$. Following the methods of ref. [16], we have

$$d\pi_t(X) = \pi_t(\mathcal{L}X)dt + \sqrt{\eta\kappa}(\pi_t(XL + L^*X) \\ - \pi_t(L + L^*)\pi_t(X))(dY_t - \sqrt{\eta}\pi_t(L + L^*)dt), \quad (3)$$

where $\mathcal{L}X = i[H_0 + u_t H_b, X] + \kappa L^*XL - \frac{\kappa}{2}(L^*LX + XL^*L)$. From the properties of $Y_t$ and the quantum Itô rule, it can be shown that the innovation process $dW_t = dY_t - \sqrt{\eta\kappa}\pi_t(L+L^*)dt$ is a quantum martingale and $dW_t^2 = dt$ [31]. Hence, it is a Wiener process from the spectral theorem [31] and the Lévy theorem [32]. A rigorous treatment of the quantum stochastic calculus and quantum filtering can be found in (e.g., [30], [31], [33]).

Note that, on the one hand, $\pi_t(X)$ is the conditional expectation of $X$ at time $t$ based on the observation of $Y_{s \leq t}$, on the other hand, the expectation of $X$ is calculated by $\text{Tr}[\rho_t X]$ in the Schrödinger picture, so we can define the conditional density operator at time $t$ by

$$\text{Tr}[\rho_t X] = \pi_t(X).$$

It is easy to get the dynamics of the conditional density operator $\rho_t$ from (3). One has

$$d\rho_t = -i[H_0 + u_t H_b, \rho_t]dt + \kappa \mathcal{D}[L]\rho_t dt + \sqrt{\eta\kappa}\mathcal{H}[L]\rho_t dW_t, \quad (4)$$

where $W_t$ is a Wiener process on some probability space $(\Omega, \mathcal{F}, \mathbb{P})$ and the superoperators $\mathcal{D}$ and $\mathcal{H}$ are defined by

$$\mathcal{D}[\Lambda]\rho = \Lambda\rho\Lambda^* - \frac{1}{2}(\Lambda^*\Lambda\rho + \rho\Lambda^*\Lambda),$$

$$\mathcal{H}[\Lambda]\rho = \Lambda\rho + \rho\Lambda^* - \text{Tr}(\Lambda\rho + \rho\Lambda^*)\rho.$$

This form of equation (4) is also known as the stochastic master equation or the quantum trajectory in physics. In this paper, we do not consider another MFC mode — direct feedback control by output measurement, see e.g., [34] and [35]. An interesting comparison between indirect feedback control (as in (4)) and the direct feedback control can be found in [36].

It is worth pointing out that in the special case where $H_0 = 0, H_b = F_y, L = F_z$, the equation (4) corresponds to the control of quantum spins, which has been investigated in depth, see e.g., ([16], [17]). While in the current paper, we will focus on more general $H_0, H_b$, and $L$ to establish some fundamental results on the capability of MFC for a wide class of quantum systems.

To be specific, our control objective is to prepare an arbitrarily desired eigenstate of $H_0$ asymptotically with probability 1 from an arbitrary initial state. Note that preparing a family of fiducial states is one of the four basic requirements for quantum computation and is the basis for subsequent manipulations, e.g., implementing a family of universal unitary operations on these fiducial states [37].

Throughout the paper, the admissible control set of MFC is defined as $\mathcal{U}$, which consists of control laws adapted to the observation process $Y_{s \leq t}$.

## III. A Selection Theorem on the Measurement Channel

Note that in contrast to the classical control theory, we should choose an appropriate measurement channel $L$ in the quantum case. This is because the measurement on a quantum state will gain some information of the state as well as introduce a state collapse in general. Remember that $L$ is called the measurement channel, because it depicts how the back action effect of the measurement on the bath is imposed on the system. In this section, we will show how to choose the measurement channel $L$ appropriately.

First, we give the following definition:

**Definition.** An eigenstate $\rho_f$ of $H_0$ is said to be ***asymptotic reachable under*** $L$ if there exit some control channel $H_b$ and control law $u_t$ such that for arbitrary initial state $\rho_0$, there exits a unique solution of (4) denoted by $\Phi_t^{(L)}(\rho_0, H_b, u)^2$, which satisfies

$$\mathbb{P}\{\lim_{t \to \infty} \Phi_t^{(L)}(\rho_0, H_b, u) = \rho_f\} = 1.$$

---

[2]In the following, it may be abbreviated as $\Phi_t(\rho_0, u)$.

Now, we only consider the case where $H_0$ and $L$ are non-degenerate[3], and we have the following theorem.

*Theorem 1:* Suppose that $H_0$ and $L$ are non-degenerate, $L = L^*$ and $[H_0, L] = 0$. Then every eigenstate of $H_0$ is asymptotic reachable under $L$.

The proof of the above theorem is based on that of Theorem 4.2 in Ref. [17].

First of all, Let us take the $N$ eigenstates $\psi_n$, $n = 1, \ldots, N$ of $H_0$ as the basis of the matrix representation of the system operators in question, where $\psi_n$ is the vector with only a nonzero element 1 in the n-th row. Then $H_0$ is a diagonal matrix. Let us define the distance between the state $\rho \in \mathcal{S}$ and one of the $H_0$ eigenstates $\rho_d = \psi_d \psi_d^*$ as

$$D(\rho, \rho_d) = 1 - \text{Tr}(\rho \rho_d).$$

Secondly, we define

$$\mathcal{S}_{>\beta} \triangleq \{\rho \in \mathcal{S} : \beta < D(\rho, \rho_d) \leq 1\},$$

$$\mathcal{S}_{\geq \beta} \triangleq \{\rho \in \mathcal{S} : \beta \leq D(\rho, \rho_d) \leq 1\},$$

$$\mathcal{S}_{\leq \beta} \triangleq \{\rho \in \mathcal{S} : 0 \leq D(\rho, \rho_d) \leq \beta\}$$

$$\mathcal{S}_{<\beta} \triangleq \{\rho \in \mathcal{S} : 0 \leq D(\rho, \rho_d) < \beta\}.$$

The proof of Theorem 1 will be proceeded by the following five lemmas whose corresponding proofs can be found in the Appendices.

*Lemma 1:* For the control model (4) with $u_t$ being a bounded and real càdlàg[4] process and $\rho_0 \in \mathcal{S}$, there exits a unique strong solution $\rho_t = \Phi_t(\rho_0, u)$ in $\mathcal{S}$.

*Lemma 2:* (Proposition 3.5 of Ref. [17]) For the control model (4) with $\rho_0 \in \mathcal{S}$ and $u_t = u(\rho_t)$ where $u \in C^1(\mathcal{S}, \mathbb{R})$, there exits a unique strong solution $\rho_t = \Phi_t(\rho_0, u)$ in $\mathcal{S}$, and $u_t$ is bounded.

---

[3]"Here non-degeneracy of a matrix means that the eigenvalues belonging to different Jordan blocks of its Jordan's normal form are mutually different" [38]. In physics, the degeneracy can be lifted by an external field.

[4]Its sample paths are right continuous and with left limits [39].

*Lemma 3:* For the control model (4) with $u_t \equiv 1$, denote $A \triangleq -i(H_0+H_b)-\kappa L^*L+\alpha\sqrt{\kappa}L$, where $\alpha$ is a real number. If there exist an $\alpha \in \mathbb{R}$ and an eigenstate $\rho_d = \psi_d\psi_d^*$ of $H_0$ such that

$$\text{rank} \begin{pmatrix} \psi_d^* \\ \psi_d^* A \\ \vdots \\ \psi_d^* A^{N-1} \end{pmatrix} = N, \tag{5}$$

then there exists $\gamma > 0$ such that for any initial state $\rho_0 \in \mathcal{S}_{>1-\gamma}$, $\rho_t$ will exit $\mathcal{S}_{>1-\gamma}$ in finite time with probability 1.

*Lemma 4:* For the feedback control model (4) with $u(\rho_t) = -\text{Tr}(i[H_b, \rho_t]\rho_d)$, denote by $\Phi_t(\rho, u)$ the solution of the model with an initial state $\rho$ and a control $u$. Then one has

$$\mathbb{P}\{\sup_{0 \le t < \infty} D(\Phi_t(\rho, u), \rho_d) \ge 1 - \gamma/2\} \le \frac{1-\gamma}{1-\gamma/2} = 1 - p < 1, \quad \text{for all } \rho \in \mathcal{S}_{\le 1-\gamma}.$$

*Lemma 5:* For the feedback control model (4) with $u(\rho_t) = -\text{Tr}(i[H_b, \rho_t]\rho_d)$, if $H_0$, $L$ are non-degenerate, $L = L^*$ and $[H_0, L] = 0$, then the trajectories of $\rho_t$ which never exit $\mathcal{S}_{<1-\gamma/2}$ will converge almost surely to $\rho_d$ as $t \to \infty$.

**Proof of Theorem 1**. We divide the proof into two steps. In the first step, we show how to select the control channel $H_b$ and to construct the corresponding control law $u_t$ according to the arbitrarily desired eigenstate of $H_0$. In the second step, we show that there exists a unique solution $\rho_t$ which converges to the desired eigenstate with probability 1 under the designed control law $u_t$.

**Step 1**. For the feedback control model (4), let us first consider how to select the control channel $H_b$ appropriately. Let $A = -i(H_0 + H_b) - \kappa L^2 + \alpha\sqrt{\kappa}L$ as defined in Lemma 3, where $\alpha$ is a real number. We proceed to select $H_b$ such that there exists a real number $\alpha$ satisfying the following rank condition:

$$\text{rank} \begin{pmatrix} \psi_d^* \\ \psi_d^* A \\ \vdots \\ \psi_d^* A^{N-1} \end{pmatrix} = N \tag{6}$$

*for all the eigenstates $\rho_d = \psi_d\psi_d^*$ $(d = 1, \cdots, N)$ of $H_0$.*

Note that this is similar to the observability condition in the linear control system theory. Now, we show how to choose the control channel $H_b$ to meet the above rank condition.

First we associate $(A; H_0, H_b, L)$ with a non-oriented graph $G(A; H_0, H_b, L) = (V, E)$, where the vertices set $V$ corresponds to the eigenstates of $H_0$, i.e., $V = \{\psi_d, d = 1, \cdots, N\}$. There is an edge between $\psi_i$ and $\psi_j$ iff $(H_b)_{i,j} \neq 0$ for $i \neq j$. Hence, $E = \{(\psi_i, \psi_j) : (H_b)_{i,j} \neq 0$ for $i \neq j\}$ does not depend on $L$.

For the feedback control model (4), a necessary condition for $A$ to meet the rank condition (6) is that the graph $G(A; H_0, H_b, L)$ is connected. Actually, Since $H_0$ is diagonal, non-degenerate and $[H_0, L] = 0$, we know that $L$ is also diagonal. Thus the off-diagonal elements of $A$ are those of $-iH_b$. Hence, if $G(A; H_0, H_b, L)$ is not connected, then we can find a permutation matrix $P$, such that [40] $A = P \begin{pmatrix} A_1 & 0 \\ 0 & A_2 \end{pmatrix} P^T$. Thus for arbitrary $d \in \{1, \cdots, N\}$ we have

$$\text{rank} \begin{pmatrix} \psi_d^* \\ \psi_d^* A \\ \vdots \\ \psi_d^* A^{N-1} \end{pmatrix} = \text{rank} \begin{pmatrix} \psi_d^* P P^T \\ \psi_d^* P \begin{pmatrix} A_1 & 0 \\ 0 & A_2 \end{pmatrix} P^T \\ \vdots \\ \psi_d^* P \begin{pmatrix} A_1^{N-1} & 0 \\ 0 & A_2^{N-1} \end{pmatrix} P^T \end{pmatrix}$$

Note that $\psi_i$, $i = 1, \cdots, N$ are the basis of the matrix representation, and $P$ is a permutation matrix, there exits some integer $k \in \{1, \cdots, N\}$, which may depend on $d$ and $P$, such that

$$\text{rank} \begin{pmatrix} \psi_d^* \\ \psi_d^* A \\ \vdots \\ \psi_d^* A^{N-1} \end{pmatrix} = \text{rank} \begin{pmatrix} \psi_k^* \\ \psi_k^* \begin{pmatrix} A_1 & 0 \\ 0 & A_2 \end{pmatrix} \\ \vdots \\ \psi_k^* \begin{pmatrix} A_1^{N-1} & 0 \\ 0 & A_2^{N-1} \end{pmatrix} \end{pmatrix} < N.$$

Therefore, the connectivity of graph $G(A; H_0, H_b, L)$ is necessary for $A$ in feedback control model (4) to meet the rank condition (6).

Intuitively, a connected graph $G(A; H_0, H_b, L)$ implies that all the eigenstates of $H_0$ (the corresponding energy levels in physical meaning) can reach each other under the control law $u_t$.

Moreover, let $G_1$ be the set of the graphs corresponding to matrices $H_b$ whose elements on the secondary diagonal are all nonzero while the others are all zero. From the proof of Lemma 4.4 in [17], it is easy to see that if $G(A; H_0, H_b, L) \in G_1$, then there exists $\alpha$ such that $A$ meets the rank condition (6). Therefore, for the feedback control model (4), a sufficient condition for $A$ to meet the rank condition (6) is to choose the control channel $H_b$ such that $G(A; H_0, H_b, L)$ is actually a path[5] of length $N - 1$.

Next, we give the feedback control law $u_t$.

In order to globally prepare an arbitrary eigenstate $\rho_d$ of $H_0$ with probability 1, let $\gamma > 0$ be as defined in Lemma 3, and define $\mathcal{B} \triangleq \{\rho : \gamma/2 < \text{Tr}(\rho\rho_d) < \gamma\}$, we construct the feedback control law as follows:

1) If $\text{Tr}(\rho_t \rho_d) \geq \gamma$, $u_t = -\text{Tr}(i[H_b, \rho_t]\rho_d)$;
2) If $\text{Tr}(\rho_t \rho_d) \leq \gamma/2$, $u_t = 1$;
3) If $\rho_t \in \mathcal{B}$, then $u_t = -\text{Tr}(i[H_b, \rho_t]\rho_d)$ if $\rho_t$ enters $\mathcal{B}$ through the boundary $\text{Tr}(\rho_t \rho_d) = \gamma$; $u_t = 1$ otherwise.

**Step 2**. In this step, we combine the results from Lemma 1 to Lemma 5 to show the existence, uniqueness and global stability of the solution of the model (4) under the control law $u_t$.

As defined in the proof of Lemma 3, $\Phi_t(\rho, u)$ denotes the solution of equation (4) at time $t$ with the initial state $\rho$ and the control $u$. Now, for an arbitrary fixed initial state $\rho \in \mathcal{S}$ and the control law $u$ given in step 1, we first construct a solution $\Phi_{t \wedge n}(\rho, u)$, where $n$ is a positive integer.

Define the stopping time

$$\tau_1^n \triangleq \tau_1 \wedge n \triangleq \inf\{t \geq 0 : \Phi_t(\rho, 1) \in \mathcal{S}_{\leq 1-\gamma}\} \wedge n.$$

Then $\Phi_{t \wedge n}(\rho, 1) = \Phi_t(\rho, 1)$, for $t < \tau_1^n$ and $\rho_{\tau_1^n} = \Phi_{\tau_1^n}(\rho, 1)$. In the following, we denote $\Phi_{s,t}(\rho_s, u)$ as the solution of equation (4) at time $t$ with the initial state $\rho_s$ at time $s$ and the control $u$, and define the stopping time

$$\sigma_1^n \triangleq \sigma_1 \wedge n \triangleq \inf\{t \geq \tau_1^n : \Phi_{\tau_1^n, t}(\rho_{\tau_1^n}, \tilde{u}) \in \mathcal{S}_{\geq 1-\gamma/2}\} \wedge n,$$

where $\tilde{u} = -\text{Tr}(i[H_b, \rho_t]\rho_d)$. Then we know for $t < \sigma_1^n$,

$$\Phi_{t \wedge n}(\rho, u) = \mathcal{X}_{t < \tau_1^n} \Phi_t(\rho, 1) + \mathcal{X}_{\tau_1^n \leq t < \sigma_1^n} \Phi_{\tau_1^n, t}(\rho_{\tau_1^n}, \tilde{u}),$$

---

[5] A path in $G(A; H_0, H_b, L)$ of length r is a sequence $[\psi_{i_0}, \cdots, \psi_{i_r}]$ of distinct vertices such that $(\psi_{i_{j-1}}, \psi_{i_j}) \in E$.

where $\mathcal{X}_A$ represents an indicator function on the set A. Then we can recursively define a sequence of stopping times as follows

$$\sigma_k^n \triangleq \sigma_k \wedge n \triangleq \inf\{t \geq \tau_k^n : \Phi_{\tau_k^n, t}(\rho_{\tau_k^n}, \tilde{u}) \in \mathcal{S}_{\geq 1-\gamma/2}\} \wedge n,$$

and

$$\tau_k^n \triangleq \tau_k \wedge n \triangleq \inf\{t \geq \sigma_{k-1}^n : \Phi_{\sigma_{k-1}^n, t}(\rho_{\sigma_{k-1}^n}, 1) \in \mathcal{S}_{\leq 1-\gamma}\} \wedge n,$$

where $\rho_{\sigma_k^n} = \Phi_{\tau_k^n, \sigma_k^n}(\rho_{\tau_k^n}, \tilde{u})$ and $\rho_{\tau_k^n} = \Phi_{\sigma_{k-1}^n, \tau_k^n}(\rho_{\sigma_{k-1}^n}, 1)$. Now we can construct the solution

$$\Phi_{t \wedge n}(\rho, u) = \mathcal{X}_{t<\tau_1^n} \Phi_t(\rho, 1) + \sum_{k=1}^{\infty} (\mathcal{X}_{\tau_k^n \leq t < \sigma_k^n} \Phi_{\tau_k^n, t}(\rho_{\tau_k^n}, \tilde{u}) + \mathcal{X}_{\sigma_k^n \leq t < \tau_{k+1}^n} \Phi_{\sigma_k^n, t}(\rho_{\sigma_k^n}, 1))$$

for $t < \sigma^n \triangleq \sigma \wedge n \triangleq \lim_{k \to \infty} \sigma_k \wedge n \leq n$. Since from Lemmas 1 and 2, every segment between any two of the stopping times is a.s. uniquely defined, it is easy to know that the solution is a.s. unique on $t \in [0, \sigma^n]$. Now let $n \to \infty$, we obtain the unique solution $\Phi_t(\rho, u)$ defined up to time $\sigma$. Furthermore, we need to prove that the solution exists for all $t \geq 0$, which can be derived just by showing $\sigma = \infty$ a.s.. Below we will show that only finitely many $\sigma_k$ are finite for almost every sample path.

We can get the fact that the strong Markov property holds on each segment between any two of the stopping times $\tau_k \leq t < \sigma_k$ and $\sigma_k \leq t < \tau_{k+1}$ from Proposition 3.7 in [17]. Hence by Lemma 4 we know that

$$\mathbb{P}\{\sigma_k < \infty | \tau_k < \infty\} \leq 1 - p,$$

and by Lemma 3 we have

$$\mathbb{P}\{\tau_k < \infty | \sigma_{k-1} < \infty\} = 1.$$

Moreover,

$$\mathbb{P}\{\tau_k < \infty | \sigma_k < \infty\} = \mathbb{P}\{\sigma_{k-1} < \infty | \tau_k < \infty\} = 1.$$

Hence we have

$$\frac{\mathbb{P}\{\sigma_k < \infty\}}{\mathbb{P}\{\sigma_{k-1} < \infty\}} = \frac{\mathbb{P}\{\tau_k < \infty | \sigma_k < \infty\} \mathbb{P}\{\sigma_k < \infty\}}{\mathbb{P}\{\tau_k < \infty\}} \frac{\mathbb{P}\{\sigma_{k-1} < \infty | \tau_k < \infty\} \mathbb{P}\{\tau_k < \infty\}}{\mathbb{P}\{\sigma_{k-1} < \infty\}}$$

$$= \mathbb{P}\{\sigma_k < \infty | \tau_k < \infty\} \mathbb{P}(\tau_k < \infty | \sigma_{k-1} < \infty)$$

$$\leq 1 - p.$$

Hence, we have

$$\mathbb{P}\{\sigma_k < \infty\} \leq (1-p)^{k-1},$$

so

$$\sum_{k=1}^{\infty} \mathbb{P}\{\sigma_k < \infty\} \leq \sum_{k=1}^{\infty} (1-p)^{k-1} = 1/p < \infty.$$

Therefore, by the Borel-Cantelli lemma [41], we know that

$$\mathbb{P}\{\lim_{k \to \infty} \sigma_k = \sigma = \infty\} = 1.$$

Thus for almost every trajectory, there exists a positive integer $K(\omega) < \infty$ such that $\sigma_k = \infty$ for any $k \geq K$, and $\sigma_k < \infty$ for any $k < K$. Then from Lemma 5, we have that $\Phi_t(\rho, u) \to \rho_d$ a.s., as $t \to \infty$. The proof of Theorem 1 is completed. $\square$

## IV. Comparison with Open-loop Control

In this section we will compare the effect of the quantum OLC and MFC in dealing with the initial state uncertainty. It is well known that in the classical control theory, the feedback control is much superior to the OLC in dealing with uncertainties, such as parameter uncertainty, structure uncertainty, external disturbance, or all, see e.g., ([26], [27]). However, the comparison of the quantum OLC and MFC is far from obvious. This is because, as we have stressed before, in the measurement-based quantum feedback control, the measurement generally introduces another kind of uncertainty on the system besides its initial uncertainty. Hence, during the control process, we have to deal with these two kinds of uncertainties. Actually, we have proposed and analyzed this question in [24] based on a special model, and demonstrated that the measurement-based quantum feedback control is still superior to the OLC in some sense.

The measurement-based quantum feedback control model is represented as equation (4). As we have mentioned, when referring to the OLC strategy, we only use the prior information to design the control law. There are two types of OLC models: (1) there is no measurement at all; (2) there is measurement but is not used to design the control law.

Our control target is also to prepare an arbitrarily desired eigenstate of $H_0$ from an arbitrary initial state.

First, we look at the case where there is no measurement. The OLC model is described by

$$\frac{d\rho_t}{dt} = -i[H_0' + u(t)F_y, \rho_t], \tag{7}$$

where $H_0'$ may be different from the effective Hamiltonian $H_0$ in equation (4) due to the measurement back action effect originally introduced there [29].

*Theorem 2:* For the OLC model (7), for arbitrary control channel $H_b$ and arbitrary control law $u_t$, an arbitrary eigenstate of $H_0$ cannot be prepared from any mixed initial state.

**Proof.** It is easy to see that the evolution of equation (7) is unitary. Consider the von Neumann entropy

$$S(\rho) = -\text{Tr}[\rho \log \rho],$$

which has the following two properties:

1) It is invariant under the unitary evolution;
2) The entropy is zero if and only if the state is pure.

Therefore, if the initial state is mixed, its entropy $S(\rho_0) > 0$, and then $S(\rho_t) \equiv S(\rho_0) > 0$ for all time $t$. Remember that the target state is an eigenstate of $H_0$, which means its entropy is 0. Hence, we cannot prepare the target state from a mixed initial state no matter how to choose the control channel $H_b$ and how to design the control law $u_t$ by using the OLC model (7). □

Note that the initial state of the system may generally be a mixed state[6] because of the inevitable interactions between the system and the bath. Hence, we cannot achieve the target no matter how to choose the control channel $H_b$ and how to design the corresponding control law $u_t$ if we use the OLC model (7).

Next, let us look at the case where there is measurement but only the prior information is used to design the control law. The corresponding OLC model is [29]

$$\frac{d\rho_t}{dt} = -i[H_0 + u_t H_b, \rho_t] + \kappa \mathcal{D}[L]\rho_t. \tag{8}$$

In contrast to equation (4), there is no diffusion part in equation (8) which depicts the measurement induced uncertainty. This is because equation (8) actually describes the evolution of the ensemble average. This form of evolution equation is usually called the master equation in physics. We have the following two theorems.

*Theorem 3:* For the OLC model (8), if $H_0$ is non-degenerate, and $[H_0, L] = 0$, then for arbitrary control channel $H_b$ and arbitrary control law $u_t$, one cannot prepare an arbitrarily desired eigenstate of $H_0$ from any mixed initial state.

---

[6]The von Neumann entropy of a mixed state is strictly greater than 0. In this sense, we say the initial state has some uncertainties.

**Proof.** Let us first consider the dynamics of $\text{Tr}(\rho_t^2)$. From equation (8), we have

$$\begin{aligned}
&\frac{d\text{Tr}(\rho_t^2)}{dt}\\
&= 2\text{Tr}[-i(H_0 + u_t H_b)\rho_t^2 + i\rho_t^2(H_0 + u_t H_b)\\
&\quad + \kappa L\rho_t L^* \rho_t - \frac{\kappa}{2}(L^*L\rho_t^2 + \rho_t^2 L^*L)]\\
&= 2\kappa \text{Tr}[L\rho_t L^*\rho_t - L^*L\rho_t^2]\\
&\leq 2\kappa \text{Tr}^{\frac{1}{2}}[L\rho_t^2 L^*]\text{Tr}^{\frac{1}{2}}[LL^*\rho_t^2] - 2\kappa\text{Tr}[L^*L\rho_t^2]\\
&\leq \kappa(\text{Tr}[L\rho_t^2 L^*] + \text{Tr}[LL^*\rho_t^2]) - 2\kappa\text{Tr}[L^*L\rho_t^2]\\
&= \kappa\text{Tr}[(LL^* - L^*L)\rho_t^2].
\end{aligned}$$

Since $H_0$ is Hermitian and non-degenerate, it is a diagonal matrix and its eigenvalues are mutually different. From this and $[H_0, L] = 0$, $L$ is also a diagonal matrix, and so $LL^* - L^*L = 0$. Hence,

$$\frac{d\text{Tr}(\rho_t^2)}{dt} \leq 0,$$

which implies that for all $t \geq 0$, $\text{Tr}(\rho_t^2) \leq \text{Tr}(\rho_0^2)$. Note that $\text{Tr}(\rho^2) = 1$ iff $\rho$ is a pure state. Hence, we conclude that the desired target cannot be prepared from a mixed initial state $\rho_0$ under the conditions of the theorem. $\square$

*Theorem 4:* For the OLC model (8), if $H_0$ is non-degenerate, and $[H_0, L] \neq 0$, then there is at least one eigenstate of $H_0$ denoted as $\rho_d$, such that no matter how to select the control channel $H_b$ and how to design the corresponding control law $u_t$, one has

$$\limsup_{t \to \infty} D(\rho_t, \rho_d) \geq \delta_d > 0,$$

where $\delta_d = \text{Tr}^2(\rho_d \mathcal{D}[L]\rho_d)/\{2[2\text{Tr}^{\frac{1}{2}}(L^*L)^2 + \text{Tr}^{\frac{1}{2}}(L^*\rho_d L)^2 + \text{Tr}^{\frac{1}{2}}(L^*LL^*L\rho_d)]^2\}$.

**Proof.** Since $H_0$ is diagonal, non-degenerate and $[H_0, L] \neq 0$, we know that there exists at least one non-diagonal entry of $L$ which is not zero. Without loss of generality, suppose it is in the $d$-th column of $L$. Hence we have

$$\text{Tr}(\rho_d \mathcal{D}[L]\rho_d) = -\sum_{i \neq d}^{N} L_{id}^* L_{id} < 0. \tag{9}$$

Next we use a contradiction argument to give the proof of this theorem. Suppose that for any eigenstate $\rho_i$ of $H_0$, their exist a corresponding control channel $H_b$ and a corresponding control

law $u_t$, such that

$$\limsup_{t\to\infty} D(\rho_t, \rho_i) < \delta_i.$$

Then there exist some $\epsilon > 0$, $T > 0$, such that for any $t > T$

$$D(\rho_t, \rho_i) < \delta_i - \epsilon.$$

Therefore for $t > T$,

$$\|\rho_t - \rho_i\| = \text{Tr}^{\frac{1}{2}}((\rho_t - \rho_i)^2) \leq (2D(\rho_t, \rho_i))^{\frac{1}{2}} < \sqrt{2(\delta_i - \epsilon)}. \tag{10}$$

Then by (10) we have

$$\begin{aligned}
&\kappa(\text{Tr}(\rho_t \mathcal{D}[L]\rho_t) - \text{Tr}(\rho_d \mathcal{D}[L]\rho_d)) \\
&= \kappa(\text{Tr}(L\rho_t L^* \rho_t) - \text{Tr}(L\rho_d L^* \rho_d) + \text{Tr}(L^* L\rho_d^2) - \text{Tr}(L^* L\rho_t^2)) \\
&= \kappa(\text{Tr}(L\rho_t L^*(\rho_t - \rho_d)) + \text{Tr}((\rho_t - \rho_d)L^* \rho_d L) + \text{Tr}(L^* L\rho_d(\rho_d - \rho_t)) + \text{Tr}(\rho_t L^* L(\rho_d - \rho_t))) \\
&\leq \kappa(\text{Tr}^{\frac{1}{2}}(L\rho_t L^*)^2 + \text{Tr}^{\frac{1}{2}}(L^* \rho_d L)^2 + \text{Tr}^{\frac{1}{2}}(L^* L L^* L\rho_d) + \text{Tr}^{\frac{1}{2}}(L^* L L^* L\rho_t^2))\|\rho_t - \rho_d\| \\
&< \sqrt{2(\delta_d - \epsilon)}\kappa(2\text{Tr}^{\frac{1}{2}}(L^* L)^2 + \text{Tr}^{\frac{1}{2}}(L^* \rho_d L)^2 + \text{Tr}^{\frac{1}{2}}(L^* L L^* L\rho_d)).
\end{aligned} \tag{11}$$

Then by (9) and (11), we have for $t > T$

$$\begin{aligned}
\frac{d\text{Tr}(\rho_t^2)}{dt} &= 2\kappa\text{Tr}(\rho_t \mathcal{D}[L]\rho_t) \\
&= 2\kappa\text{Tr}(\rho_d \mathcal{D}[L]\rho_d) + 2\kappa(\text{Tr}(\rho_t \mathcal{D}[L]\rho_t) - \text{Tr}(\rho_d \mathcal{D}[L]\rho_d)) \\
&< 2\kappa\text{Tr}(\rho_d \mathcal{D}[L]\rho_d) + 2\sqrt{2(\delta_d - \epsilon)}\kappa(2\text{Tr}^{\frac{1}{2}}(L^* L)^2 + \text{Tr}^{\frac{1}{2}}(L^* \rho_d L)^2 + \text{Tr}^{\frac{1}{2}}(L^* L L^* L\rho_d)) \\
&\triangleq \alpha(\epsilon).
\end{aligned}$$

By substituting the definition of $\delta_d$ into $\alpha(\epsilon)$ defined above we can find that $\alpha(\epsilon) < 0$. Hence, from the above inequality, we have a contradiction

$$0 \leq \text{Tr}(\rho_t^2) \to -\infty, \text{ as } t \to \infty,$$

and the proof is completed. $\square$

From Theorems 2, 3 and Theorem 4, we know that if using the OLC model, we cannot achieve an arbitrarily desired target state no matter how to select the measurement channel $L$, the control channel $H_b$ and the control law $u_t$. Hence, in comparison with the MFC results as established in Theorem 1, we conclude that the measurement-based quantum feedback control is still superior to the quantum OLC in dealing with the initial state uncertainty.

Note that in contrast to the classical control theory, the measurement-based quantum feedback control model (4) is different from the OLC models (7) and (8). This is due to the inherent quantum measurement back action effect which consists of the deterministic drift part and the uncertainty part. Specifically, the change from $H_0'$ in (7) to $H_0$ in (4) and $\kappa \mathcal{D}[L]\rho_t dt$ describe the deterministic back action effect, while $\sqrt{\kappa\eta}\mathcal{H}[L]\rho_t dW_t$ depicts the uncertainty part. It is these structural changes, especially the uncertainty part, that make it possible to design a feedback control law to achieve the control target.

## V. THE IMPOSSIBILITY THEOREM OF MFC

In this section, we will give an impossibility theorem on the MFC if the selected measurement channel $L$ and the effective Hamiltonian $H_0$ do not commute.

*Theorem 5:* For the MFC model (4), if $H_0$ is non-degenerate, $[H_0, L] \neq 0$ and $\eta \in [0, 1)$, then there is at least one eigenstate of $H_0$ denoted as $\rho_d$, such that no matter how to select the control channel $H_b$ and how to design the corresponding control law $u_t$, one has

$$\limsup_{t\to\infty} D(\rho_t, \rho_d) \geq \Delta_d > 0, \quad a.s.$$

where $\Delta_d = [2\mathrm{Tr}(\rho_d \mathcal{D}[L]\rho_d) + \eta \mathrm{Tr}(\mathcal{H}[L]\rho_d)^2]^2 / \{2[2\varphi_1(L, \rho_d) + \eta\varphi_2(L, \rho_d)]^2\}$,

$$\varphi_1(L, \rho_d) = 2\mathrm{Tr}^{\frac{1}{2}}(L^*L)^2 + \mathrm{Tr}^{\frac{1}{2}}(L^*\rho_d L)^2 + \mathrm{Tr}^{\frac{1}{2}}(L^*LL^*L\rho_d),$$

$$\varphi_2(L, \rho_d) = 2\mathrm{Tr}^{\frac{1}{2}}(L^*L)^2 + 2\mathrm{Tr}^{\frac{1}{2}}(L^*LL^*L\rho_d) + 2\mathrm{Tr}(L^*L) + 2\mathrm{Tr}^{\frac{1}{2}}(LL^*\rho_d L^*L\rho_d)$$
$$+ 3\mathrm{Tr}^{\frac{1}{2}}(L + L^*)^2 \mathrm{Tr}((L + L^*)\rho_d) + 3\mathrm{Tr}^{\frac{1}{2}}(L + L^*)^2 \mathrm{Tr}^{\frac{1}{2}}(L + L^*)^2$$
$$+ 2\mathrm{Tr}^{\frac{1}{2}}(L + L^*)^2 \mathrm{Tr}^{\frac{1}{2}}((L + L^*)^2 \rho_d).$$

**Proof.** The proof of this theorem is similar to the proof of Theorem 4 and a contradiction argument will be used. First, since $H_0$ is diagonal, non-degenerate and $[H_0, L] \neq 0$, we know that $L$ has at least one non-zero non-diagonal entry. For simplicity, we suppose it is in the $d$-th column of $L$. Thus for the eigenstate $\rho_d$, we have

$$2\kappa \mathrm{Tr}(\rho_d \mathcal{D}[L]\rho_d) + \eta\kappa \mathrm{Tr}(\mathcal{H}[L]\rho_d)^2 = -2\kappa(1-\eta)\sum_{i\neq d}^{N} L_{id}^* L_{id} \triangleq -2\kappa(1-\eta)\alpha < 0. \quad (12)$$

Suppose that for any eigenstate $\rho_i$ of $H_0$, there exist corresponding control channel $H_b$ and control law $u_t$, such that

$$\limsup_{t\to\infty} D(\rho_t, \rho_i) < \Delta_i, \quad \text{on } A,$$

where $A$ is a set with positive probability $\mathbb{P}\{A\} > 0$. Then for each sample point $\omega \in A$, there exist $\epsilon > 0$ and $T(\omega) > 0$, such that for any $t > T(\omega)$ we have

$$D(\rho_t(\omega), \rho_i) < \Delta_i - \epsilon.$$

Therefore, for $t > T(\omega)$ with $\omega \in A$, we have

$$\|\rho_t(\omega) - \rho_i\|^2 \leq 2D(\rho_t(\omega), \rho_i) < 2(\Delta_i - \epsilon). \tag{13}$$

Then for each $\omega \in A$, by (13) we have

$$\begin{aligned}
&\kappa(\text{Tr}(\rho_t \mathcal{D}[L]\rho_t) - \text{Tr}(\rho_d \mathcal{D}[L]\rho_d)) \\
&= \kappa(\text{Tr}(L\rho_t L^* \rho_t) - \text{Tr}(L\rho_d L^* \rho_d) + \text{Tr}(L^* L \rho_d^2) - \text{Tr}(L^* L \rho_t^2)) \\
&= \kappa(\text{Tr}(L\rho_t L^*(\rho_t - \rho_d)) + \text{Tr}((\rho_t - \rho_d)L^* \rho_d L) + \text{Tr}(L^* L \rho_d (\rho_d - \rho_t)) + \text{Tr}(\rho_t L^* L(\rho_d - \rho_t))) \\
&\leq \kappa(\text{Tr}^{\frac{1}{2}}(L\rho_t L^*)^2 + \text{Tr}^{\frac{1}{2}}(L^* \rho_d L)^2 + \text{Tr}^{\frac{1}{2}}(L^* L L^* L \rho_d) + \text{Tr}^{\frac{1}{2}}(L^* L L^* L \rho_t^2))\|\rho_t - \rho_d\| \\
&< \sqrt{2(\Delta_d - \epsilon)}\kappa(2\text{Tr}^{\frac{1}{2}}(L^* L)^2 + \text{Tr}^{\frac{1}{2}}(L^* \rho_d L)^2 + \text{Tr}^{\frac{1}{2}}(L^* L L^* L \rho_d)) \\
&= \kappa \varphi_1(L, \rho_d) \sqrt{2(\Delta_d - \epsilon)},
\end{aligned} \tag{14}$$

and similarly for $\omega \in A$

$$\begin{aligned}
&\text{Tr}(\mathcal{H}[L]\rho_t)^2 - \text{Tr}(\mathcal{H}[L]\rho_d)^2 \\
&= 2\text{Tr}(L^* L \rho_t^2) - 2\text{Tr}(L^* L \rho_d^2) + \text{Tr}(L\rho_t L\rho_t) - \text{Tr}(L\rho_d L\rho_d) + \text{Tr}(L^* \rho_t L^* \rho_t) - \text{Tr}(L^* \rho_d L^* \rho_d) \\
&\quad - 2\text{Tr}((L + L^*)\rho_t)\text{Tr}((L + L^*)\rho_t^2) + 2\text{Tr}((L + L^*)\rho_d)\text{Tr}((L + L^*)\rho_d^2) \\
&\quad + \text{Tr}^2((L + L^*)\rho_t)\text{Tr}(\rho_t^2) - \text{Tr}^2((L + L^*)\rho_d)\text{Tr}(\rho_d^2) \\
&< 2(\text{Tr}^{\frac{1}{2}}(L^* L)^2 + \text{Tr}^{\frac{1}{2}}(L^* L L^* L \rho_d))\|\rho_t - \rho_d\| + \text{Tr}(L\rho_t L(\rho_t - \rho_d)) + \text{Tr}((\rho_t - \rho_d)L\rho_d L) \\
&\quad + \text{Tr}(L^* \rho_t L^*(\rho_t - \rho_d)) + \text{Tr}((\rho_t - \rho_d)L^* \rho_d L^*) + 2\text{Tr}((L + L^*)\rho_d)\text{Tr}((L + L^*)(\rho_d - \rho_t)) \\
&\quad + 2\text{Tr}((L + L^*)\rho_t)\text{Tr}((L + L^*)\rho_d(\rho_d - \rho_t)) + 2\text{Tr}((L + L^*)\rho_t)\text{Tr}((L + L^*)(\rho_d - \rho_t)\rho_t) \\
&\quad + \text{Tr}((L + L^*)(\rho_d + \rho_t))\text{Tr}((L + L^*)(\rho_t - \rho_d)) \\
&< \{2\text{Tr}^{\frac{1}{2}}(L^* L)^2 + 2\text{Tr}^{\frac{1}{2}}(L^* L L^* L \rho_d) + 2\text{Tr}(L^* L) + 2\text{Tr}^{\frac{1}{2}}(L L^* \rho_d L^* L \rho_d) \\
&\quad + 3\text{Tr}^{\frac{1}{2}}(L + L^*)^2 \text{Tr}((L + L^*)\rho_d) + 3\text{Tr}^{\frac{1}{2}}(L + L^*)^2 \text{Tr}^{\frac{1}{2}}(L + L^*)^2 \\
&\quad + 2\text{Tr}^{\frac{1}{2}}(L + L^*)^2 \text{Tr}^{\frac{1}{2}}((L + L^*)^2 \rho_d)\}\|\rho_t - \rho_d\| \\
&= \varphi_2(L, \rho_d)\|\rho_t - \rho_d\| < \varphi_2(L, \rho_d)\sqrt{2(\Delta_d - \epsilon)}.
\end{aligned} \tag{15}$$

By the Itô formula, we have

$$dTr(\rho_t^2) = 2\kappa Tr(\rho_t \mathcal{D}[L]\rho_t)dt + \eta\kappa Tr(\mathcal{H}[L]\rho_t)^2 dt + 2\sqrt{\eta\kappa}Tr(\rho_t\mathcal{H}[L]\rho_t)dW_t$$
$$= (2\kappa Tr(\rho_d \mathcal{D}[L]\rho_d) + \eta\kappa Tr(\mathcal{H}[L]\rho_d)^2)dt + (2\kappa Tr(\rho_t\mathcal{D}[L]\rho_t) + \eta\kappa Tr(\mathcal{H}[L]\rho_t)^2$$
$$- 2\kappa Tr(\rho_d\mathcal{D}[L]\rho_d) - \eta\kappa Tr(\mathcal{H}[L]\rho_d)^2)dt + 2\sqrt{\eta\kappa}Tr(\rho_t\mathcal{H}[L]\rho_t)dW_t. \tag{16}$$

Then by (12), (14), (15) and (16), for $t_0 > T(\omega)$ with $\omega \in A$, we have

$$Tr(\rho_t^2) = Tr(\rho_{t_0}^2) + \int_{t_0}^{t} 2\kappa Tr(\rho_d\mathcal{D}[L]\rho_d) + \eta\kappa Tr(\mathcal{H}[L]\rho_d)^2 ds$$
$$+ \int_{t_0}^{t} 2\kappa Tr(\rho_s\mathcal{D}[L]\rho_s) - 2\kappa Tr(\rho_d\mathcal{D}[L]\rho_d) - \eta\kappa Tr(\mathcal{H}[L]\rho_d)^2 + \eta\kappa Tr(\mathcal{H}[L]\rho_s)^2 ds$$
$$+ 2\sqrt{\eta\kappa}\int_{t_0}^{t} Tr(\rho_s\mathcal{H}[L]\rho_s)dW_s$$
$$< Tr(\rho_{t_0}^2) + \{-2\kappa(1-\eta)\alpha + 2\kappa\varphi_1(L,\rho_d)\sqrt{2(\Delta_d - \epsilon)}$$
$$+ \eta\kappa\varphi_2(L,\rho_d)\sqrt{2(\Delta_d - \epsilon)}\}(t - t_0) + 2\sqrt{\eta\kappa}\int_{t_0}^{t} Tr(\rho_s\mathcal{H}[L]\rho_s)dW_s. \tag{17}$$

Moreover, by Theorem 1.51 in [42], for $0 < \epsilon < \frac{1}{2}$, we have

$$Tr(\rho_t^2) < Tr(\rho_{t_0}^2) + \{-2\kappa(1-\eta)\alpha + 2\kappa\varphi_1(L,\rho_d)\sqrt{2(\Delta_d - \epsilon)}$$
$$+ \eta\kappa\varphi_2(L,\rho_d)\sqrt{2(\Delta_d - \epsilon)}\}(t - t_0) + o(t^{\frac{1}{2}+\epsilon}) + O(1).$$
$$\to -\infty, \quad \text{as } t \to \infty, \text{on } A$$

where the limit is derived by the definition of $\Delta_d$. This is a contradiction. Thus the theorem is proved. $\square$

Note that if $[H_0, L] = 0$, then all the off-diagonal elements of $L$ must be zero, so by eq. (12), we have $\Delta_d = 0$. Hence, the above limit to the MFC is due to the non-perfect detection and the noncommutative relationship between the effective Hamiltonian $H_0$ and the measurement channel $L$. Theorem 5 is actually inspired by the Heisenberg uncertainty principle which says that given two observable $H_0$ and $L$, for any quantum state $\rho$, their standard deviations satisfy

$$\Delta H_0 \cdot \Delta L \geq \frac{|\langle [H_0, L] \rangle|}{2},$$

where the mean value of $H_0$ is $\langle H_0 \rangle = Tr[H_0\rho]$ and the standard deviation of $H_0$ is $\Delta H_0 = \sqrt{\langle H_0^2 \rangle - \langle H_0 \rangle^2}$. We see that the non-commutability plays an important role in the Heisenberg uncertainty principle, just as the role played in the asymptotic reachability of Theorem 5.

Combining Theorem 1 and Theorem 5, one has

*Theorem 6:* Suppose $H_0$ and $L$ are non-degenerate, $L = L^*$, and $\eta \in (0, 1)$. Then every eigenstate of $H_0$ is asymptotic reachable under $L$ if and only if $[H_0, L] = 0$.

## VI. CONCLUDING REMARKS

In this paper, three fundamental problems concerning the asymptotic reachability of eigenstates by the MFC are investigated. First, we have given a selection theorem on how to select the measurement channel. Secondly, we have compared the MFC and the OLC in dealing with the system initial uncertainty, and have shown that the MFC is still superior to the OLC for the quantum systems in spite of the additional uncertainty induced by the measurement. Thirdly, we have given an impossibility theorem on the capability of the MFC. It is worth emphasizing that the study on the capability of MFC is just initiated. For future investigations, it may be necessary to further study the limitations of quantum MFC, e.g., in dealing with structural uncertainties and/or in achieving other control objectives. Also, it would be interesting to study how the measurement channel and/or the control channel can be adjusted adaptively according to the quantum state in question, as discussed in e.g., [44]. The study of the above problems with the non-Markovian model would also be interesting. There is no doubt that these investigations will help us understand more about the MFC.

## VII. APPENDICES

In this appendix, we will give the proofs of Lemmas 1, 3, 4 and 5. The analyzes are mainly based on those in [17] but for the more general model in the paper.

### A. Support Theorem

The following support theorem from e.g. ([17], [39]) connects the properties of solutions of a stochastic differential equation to solutions of an associated deterministic system.

*Theorem 7:* Let M be a connected, paracompact $C^\infty$-manifold and let $X_k, k = 0, ..., n$, be $C^\infty$ vector fields on M such that all linear sums of $X_k$ are complete. Let $X_k = \sum_l X_k^l(x) \partial_l$ in local coordinates and consider the Stratonovich equation

$$dx_t = X_0(x_t) + \sum_{k=1}^n X_k(x_t) \circ dW_t^k, \quad x_0 = x.$$

Consider in addition the associated deterministic control system

$$\frac{d}{dt} x_t^u = X_0(x_t^u) + \sum_{k=1}^n X_k(x_t^u) \circ u^k(t), \quad x_0^u = x,$$

with $u^k \in \mathscr{U}$, the set of all piecewise constant functions from $\mathcal{R}_+$ to $\mathcal{R}$. Then

$$\mathscr{S}_x = \overline{\{x_\cdot^u : u \in \mathscr{U}^n\}} \subset \mathscr{W}_x,$$

where $\mathscr{W}_x$ is the set of all continuous paths from $\mathcal{R}_+$ to $M$ starting at $x$, equipped with the topology of uniform convergence on compact sets, and $\mathscr{S}_x$ is the smallest closed subset of $\mathscr{W}_x$ such that $\mathbb{P}\{\omega \in \Omega : x_\cdot(\omega) \in \mathscr{S}_x\} = 1$.

## B. Proof of Lemma 1

**Proof**. Consider the corresponding unnormalized linear form of equation (4)

$$d\tilde{\rho}_t = -i[H_0 + u_t H_b,\ \tilde{\rho}_t]dt + \kappa(L\tilde{\rho}_t L^* - \frac{1}{2}(L^*L\tilde{\rho}_t + \tilde{\rho}_t LL^*))dt + \sqrt{\eta\kappa}(L\tilde{\rho}_t + \tilde{\rho}_t L^*)dY_t. \quad (18)$$

By [39] (pp.195-197), it is not difficult to check that the above equation has a unique strong solution as it obeys a functional Lipschitz condition.

Now we introduce a process $\bar{W}_t = \sqrt{\eta} Y_t + \sqrt{1-\eta}\hat{W}$, where $\hat{W}_t$ is a Wiener process which is independent with $Y_t$. Then with the same initial condition $\tilde{\rho}_0 = \bar{\rho}_0(= \rho_0)$, the process $\tilde{\rho}_t$ is stochastically equivalent to $\mathbb{E}[\bar{\rho}_t | \mathcal{F}_t^Y]$, where $\bar{\rho}_t$ satisfies the following equation

$$d\bar{\rho}_t = -i[H_0 + u_t H_b,\ \bar{\rho}_t]dt + \kappa(L\bar{\rho}_t L^* - \frac{1}{2}(L^*L\bar{\rho}_t + \bar{\rho}_t LL^*))dt + \sqrt{\kappa}(L\bar{\rho}_t + \bar{\rho}_t L^*)d\bar{W}_t. \quad (19)$$

If $\bar{\rho}_0 = \psi_0 \psi_0^*$ is a pure state, it is easy to check by the Itô formula that the solution of the equation

$$d\bar{\psi}_t = -i(H_0 + u_t H_b)\bar{\psi}_t dt - \frac{\kappa}{2}L^*L\bar{\psi}_t dt + \sqrt{\kappa}L\bar{\psi}_t d\bar{W}_t \quad (20)$$

satisfies $\bar{\rho}_t = \bar{\psi}_t \bar{\psi}_t^*$. From [39], pp. 326 we know that $\bar{\psi}_t \neq 0$ a.s. for $\psi_0 \neq 0$. Therefore, $\tilde{\rho}_t = \mathbb{E}[\bar{\rho}_t | \mathcal{F}_t^Y] > 0$ a.s.. Note that equation (18) and (19) are both linear, hence, if the initial state $\rho_0$ is mixed, i.e., $\rho_0 = \sum_i \lambda_i \psi_0^i \psi_0^{i*}$ with the convex weights $\lambda_i > 0$, we can also get $\tilde{\rho}_t = \mathbb{E}[\bar{\rho}_t | \mathcal{F}_t^Y] > 0$ a.s.. This implies that $\text{Tr}\tilde{\rho}_t > 0$ a.s.. Now it is easy to prove that equation (4) with $u_t \in \mathcal{U}$ has a unique solution $\rho_t = \tilde{\rho}_t / \text{Tr}(\tilde{\rho}_t) \in \mathcal{S}$ by the Itô formula. $\square$

## C. Proof of Lemma 3

**Proof**. First of all, from Lemma 1, we know that equation (4) with $u_t \equiv 1$ has a unique solution in the compact set $\mathcal{S}$. Now for some fixed $\rho_d = \psi_d \psi_d^*$, we define the function

$$g(\rho) = \min_{t \in [0,T]}\ \mathbb{E}\ D(\Phi_t(\rho, 1), \rho_d), \quad \rho \in \mathcal{S},$$

where $T$ is a fixed time and $\Phi_t(\rho, 1)$ denotes the solution of equation (4) at time $t$ with initial state $\rho_0 = \rho$ and control $u_t \equiv 1$. The proof of Lemma 3 is divided into three steps.

**Step 1**. We first show that $g(\rho) < 1$ for any $\rho \in \mathcal{S}_1 \triangleq \{\rho \in \mathcal{S} : D(\rho, \rho_d) = 1\}$.

A contradiction argument is now used to complete the proof. Suppose that $g(\rho) = 1$ for some $\rho \in \mathcal{S}_1$, i.e., $\mathbb{E}\mathrm{Tr}(\Phi_t(\rho, 1)\rho_d) = 0$, for some $\rho \in \mathcal{S}_1$ and all $t \in [0, T]$.

From the proof of Lemma 1, we can write $\Phi_t(\rho, 1) = \tilde{\rho}_t / \mathrm{Tr}(\tilde{\rho}_t)$ with $\tilde{\rho}_t = \sum_i \lambda_i \mathbb{E}[\bar{\psi}_t^i \bar{\psi}_t^{i*} | \mathcal{F}_t^Y]$, where $\lambda_i$'s are the coefficients of the spectral decomposition of $\rho = \rho_0 = \sum_i \lambda_i \psi_0^i \psi_0^{i*}$, $\psi_0^i \in \mathcal{R}^N \setminus \{0\}$ and $\bar{\psi}_t^i$ obey the equation

$$d\bar{\psi}_t^i = -i(H_0 + H_b)\bar{\psi}_t^i dt - \frac{\kappa}{2} L^* L \bar{\psi}_t^i dt + \sqrt{\kappa} L \bar{\psi}_t^i d\bar{W}_t, \tag{21}$$
$$\bar{\psi}_0^i = \psi_0^i.$$

Now, one can show that $\mathbb{E}\mathrm{Tr}(\Phi_t(\rho, 1)\rho_d) = 0$ is equivalent to

$$\mathbb{E}\mathrm{Tr}(\tilde{\rho}_t \rho_d) = \sum_i \lambda_i \mathbb{E}[(\psi_d^* \bar{\psi}_t^i)^* (\psi_d^* \bar{\psi}_t^i)] = 0,$$

i.e., $g(\rho) = 1$ iff $\mathbb{P}\{\psi_d^* \bar{\psi}_t^i = 0\} = 1$, for all $i$ and $t \in [0, T]$.

Remember that

$$d\bar{W}_t = \sqrt{\eta} dY_t + \sqrt{1-\eta} d\hat{W}_t$$
$$= \eta\sqrt{\kappa}\mathrm{Tr}[(L + L^*)\rho_t]dt + \sqrt{\eta}dW_t + \sqrt{1-\eta}d\hat{W}_t,$$

where $W_t$ and $\hat{W}_t$ are independent Wiener processes. Since $\mathrm{Tr}[(L + L^*)\rho_t]$ is bounded, by Girsanov Theorem [32], there exists a measure $\mathbb{Q}$, such that the stochastic process $\bar{W}_t$ is a Wiener process in $[0, T]$ under the new measure $\mathbb{Q}$ and $\mathbb{Q}$ is equivalent to the measure $\mathbb{P}$. Hence,

$$g(\rho) = 1 \quad \text{iff} \quad \mathbb{Q}\{\psi_d^* \bar{\psi}_t^i = 0\} = 1, \quad \text{for all } i \text{ and } t \in [0, T]. \tag{22}$$

Moreover, it is easy to find that the linear Stratonovich form of equation (21) is

$$d\bar{\psi}_t^i = -i(H_0 + H_b)\bar{\psi}_t^i dt - \kappa L^* L \bar{\psi}_t^i dt + \sqrt{\kappa} L \bar{\psi}_t^i \circ d\bar{W}_t, \tag{23}$$
$$\bar{\psi}_0^i = \psi_0^i.$$

Now we apply Theorem 7. Recall that $\mathscr{W}_{\bar{\psi}_0}^i$ denotes the set of all continuous paths starting at $\bar{\psi}_0^i$ and $\mathscr{S}_{\bar{\psi}_0}^i$ denotes the smallest closed subset of $\mathscr{W}_{\bar{\psi}_0}^i$ satisfying that $\mathbb{Q}\{\omega \in \Omega : \bar{\psi}_t^i(\omega) \in \mathscr{S}_{\bar{\psi}_0}^i\} = 1$. Moreover, we denote $\mathscr{S}_{\bar{\psi}_0,t}^i \triangleq \{v_t \in \mathscr{W}_{\bar{\psi}_0}^i : v_t^* \psi_d = 0\}$. It is easy to prove that $\mathscr{S}_{\bar{\psi}_0,t}^i$ is closed in the compact uniform topology for any $t \in [0, T]$. Then it is obvious that $\mathscr{S}^i_{\bar{\psi}_0} \subset \mathscr{S}^i_{\bar{\psi}_0, t}$ for all $i$ and $t \in [0, T]$ from (22).

Now consider the following deterministic differential equation

$$\frac{d\psi^i_t}{dt} = A\psi^i_t = (-i(H_0 + H_b) - \kappa L^*L + \alpha\sqrt{\kappa}L)\psi^i_t, \quad \psi^i_0 = \bar{\psi}_0^{\,i}. \tag{24}$$

Then by Theorem 7 we know that the solution $\psi^i_t$ of equation (24) with some $\alpha$ satisfying equation (5) belongs to $\mathscr{S}^i_{\bar{\psi}_0}$, which implies $\psi^i_t \in \mathscr{S}_{\bar{\psi}_0, t}$, for all $t \in [0, T]$. That is to say,

$$\psi^{i*}_t \psi_d = 0, \text{ for all } t \in [0, T]. \tag{25}$$

However, from equation (24) and (25) we have

$$\frac{d^n}{dt^n}\psi^{i*}_t \psi_d|_{t=0} = (\psi^*_d A^n)\psi^i_0 = 0, \quad n = 1, 2, \ldots N-1,$$

hence,

$$\begin{pmatrix} \psi^*_d \\ \psi^*_d A \\ \vdots \\ \psi^*_d A^{N-1} \end{pmatrix} \psi^i_0 = 0.$$

Consequently, from (5) we know that $\psi^i_0 = 0$, a contradiction, and the proof of step 1 is completed.

**Step 2**. We show that there exists $\gamma > 0$ such that $g(\rho) < 1 - \gamma$ for all $\rho \in \mathcal{S}_{\geq 1-\gamma}$.

We first prove that there exists $x > 0$ such that $g(\rho) \leq 1 - x$ for all $\rho \in \mathcal{S}_{>1-x}$. Otherwise, suppose that for any $x > 0$ there exists $\rho_x \in \mathcal{S}_{>1-x}$ such that $1 - x < g(\rho_x) \leq 1$.

Take a sequence $x_n \downarrow 0$ such that the corresponding sequence $\rho_{x_n} \in \mathcal{S}_{>1-x_n}$ and $g(\rho_{x_n}) \to 1$. Because of the compactness of $\mathcal{S}$, we can take a subsequence denoted also as $x_n$ such that $\rho_{x_n} \to \rho_\infty \in \mathcal{S}_1$. From step 1 we have $g(\rho_\infty) < 1$, i.e., there exist $s \in [0, T]$ and $\epsilon > 0$ such that

$$\mathbb{E}D(\Phi_s(\rho_\infty, 1), \rho_d) = 1 - \epsilon.$$

Because the solution of equation (4) is Feller continuous by Proposition 3.6 in [17], we have

$$1 = \lim_{n \to \infty} g(\rho_{x_n}) \leq \lim_{n \to \infty} \mathbb{E}D(\Phi_s(\rho_{x_n}, 1), \rho_d) = \mathbb{E}D(\Phi_s(\rho_\infty, 1), \rho_d) = 1 - \epsilon < 1,$$

which is a contraction. Hence, there exists $x > 0$ such that $g(\rho) \leq 1 - x$ for all $\rho \in \mathcal{S}_{>1-x}$. Take $\gamma = x/2$, then the proof of step 2 is completed.

**Step 3**. In this step, we will give the proof of this lemma. Let $\tau_\rho$ be the first exit time of $\Phi_t(\rho, 1)$ from $\mathcal{S}_{>1-\gamma}$ with the initial state $\rho \in \mathcal{S}_{>1-\gamma}$. To complete the proof of this lemma, we just need to show that

$$\sup_{\rho \in \mathcal{S}_{>1-\gamma}} \mathbb{E}\tau_\rho < \infty. \tag{26}$$

Actually, from Lemma 4.3 in [43] we have

$$\mathbb{E}\tau_\rho \leq \frac{T}{1 - \sup_{\varrho \in \mathcal{S}} \mathbb{P}\{\tau_\varrho > T\}}.$$

Then in order to prove (26), we just need to show that

$$\sup_{\varrho \in \mathcal{S}} \mathbb{P}\{\tau_\varrho > T\} < 1.$$

Obviously the inequality holds if $\varrho \in \mathcal{S} \setminus \mathcal{S}_{>1-\gamma}$. Let us just consider $\varrho \in \mathcal{S}_{>1-\gamma}$. We use a contradiction argument. Suppose that for every $\epsilon > 0$, there exists $\varrho(\epsilon) \in \mathcal{S}_{>1-\gamma}$, such that $\mathbb{P}\{\tau_{\varrho(\epsilon)} > T\} > 1 - \epsilon$. Then we can get

$$\mathbb{E}D(\Phi_s(\varrho(\epsilon), 1), \rho_d) > (1-\gamma)(1-\epsilon), \quad \text{for all } s \in [0, T].$$

Now we take a sequence $\epsilon_n \downarrow 0$, then we can find a corresponding subsequence denoted also as $\varrho(\epsilon_n)$ such that $\varrho(\epsilon_n) \downarrow \varrho(\infty) \in \mathcal{S}_{\geq 1-\gamma}$ as $n \to \infty$ by the compactness of $\mathcal{S}$. Hence by the Feller continuity of the solution of equation (4), we have

$$\mathbb{E}D(\Phi_s(\varrho(\infty), 1), \rho_d) \geq (1-\gamma), \quad \text{for all } s \in [0, T],$$

which is in contradiction with the result of Step 2. Hence the inequality (26) holds, and the proof of this lemma is completed. □

*D. Proof of Lemma 4*

**Proof**. From Lemma 2, we know that equation (4) with $u(\rho_t) = -\text{Tr}(i[H_0, \rho_t]\rho_d)$ has a unique solution in the compact set $\mathcal{S}$. By the Itô formula, we can get

$$dD(\rho_t, \rho_d) = -u_t^2 dt - \sqrt{\eta\kappa}\text{Tr}[(L\rho_t + \rho_t L^* - \text{Tr}[(L+L^*)\rho_t]\rho_t)\rho_d]dW_t. \tag{27}$$

Denote $\mathcal{L}$ as the the weak infinitesimal operator of $\Phi_t(\rho, u)$, then from (27) we have

$$\mathcal{L}D(\Phi_0(\rho, u), \rho_d) = -u^2(\rho) \leq 0,$$

from which the proof can be completed by Theorem 2.2 in [17]. □

*E. Proof of Lemma 5*

**Proof**. First of all, from Lemma 2, we know that equation (4) with $u(\rho_t) = -\text{Tr}(i[H_0, \rho_t]\rho_d)$ has a unique solution in the compact set $\mathcal{S}$. We divide the proof into two steps. In the first step, we will show that the trajectories of $\rho_t$ which never exit $\mathcal{S}_{<1-\gamma/2}$ converge in probability to $\rho_d$. Then we will complete the proof of this lemma in the second step.

**Step 1**. Consider the following Lyapunov function

$$V(\rho) = 1 - \text{Tr}^2(\rho \rho_d).$$

It is easy to see that $V(\rho) \geq 0$ in $\mathcal{S}$ and $V(\rho) = 0$ iff $\rho = \rho_d$.

Note that $H_0$ is diagonal, non-degenerate and $[H_0, L] = 0$, then we know that $L$ is also diagonal. In addition, $L$ is non-degenerate, so we know that $L$ and $H_0$ have the same eigenstates. Then from the Itô formula,

$$\begin{aligned}dV(\rho_t) = &(-2u_t^2 \text{Tr}(\rho_t \rho_d) - 4\eta\kappa(\lambda_d - \text{Tr}(L\rho_t))^2 \text{Tr}^2(\rho \rho_d))dt \\ &- 2\sqrt{\eta\kappa}\text{Tr}((L\rho_t + \rho_t L^* - \text{Tr}[(L+L^*)\rho_t]\rho_t)\rho_d)\text{Tr}(\rho_t \rho_d)dW_t.\end{aligned} \tag{28}$$

Denote by $\mathcal{L}$ the weak infinitesimal operator of $\rho_t$, then from (28) we have

$$\mathcal{L}V(\rho) = -2u(\rho)^2 \text{Tr}(\rho \rho_d) - 4\eta\kappa(\lambda_d - \text{Tr}(L\rho))^2 \text{Tr}^2(\rho \rho_d) \leq 0, \tag{29}$$

where $\lambda_d$ is the eigenvalue of $L$ corresponding to the eigenstate $\rho_d$. Hence from Proposition 3.4 and 3.7 in [17] and the stochastic stability theory (Theorem 2.3 in [17]), we know that $\rho_t$ converges in probability to the largest invariant set contained in $\mathcal{N}_\mathcal{S} = \{\rho \in \mathcal{S} : \mathcal{L}V(\rho) = 0\}$. Since we just consider the trajectories that never exit $\mathcal{S}_{<1-\gamma/2}$, from (29) we know that $\mathcal{N}_\mathcal{S} \cap \mathcal{S}_{<1-\gamma/2}$ must be a subset of the largest invariant set $\mathcal{N} = \{\rho \in \mathcal{S} : \text{Tr}(L\rho) = \lambda_d\}$. Moreover, the invariant set $\mathcal{N}$ can only contain $\rho$ such that $\text{Tr}(\Phi_t(\rho, u)L)$ is constant. Therefore, by the Itô formula,

$$0 = d\text{Tr}(L\rho_t) = -iu_t\text{Tr}([H_b, \rho_t]L)dt + 2\sqrt{\eta\kappa}(\text{Tr}(L^2\rho_t) - \text{Tr}^2(L\rho_t))dW_t.$$

Hence we must have

$$\text{Tr}(L^2\rho) - \text{Tr}^2(L\rho) = 0.$$

Since $L$ is non-degenerate, we conclude that $\rho$ must be the eigenstate of $L$. Therefore, it is easy to see that $\mathcal{N}_\mathcal{S} \cap \mathcal{S}_{<1-\gamma/2} = \{\rho_d\}$, so we know that the trajectories that never exit $\mathcal{S}_{<1-\gamma/2}$ must converge in probability to $\rho_d$.

**Step 2**. In this step, we will give the proof of this lemma. Let us define the even $B = \{\omega \in \Omega : \rho_t \text{ never exits } \mathcal{S}_{<1-\gamma/2}\}$. Then from step 1 we have

$$\lim_{t\to\infty} \mathbb{P}\{D(\rho_t, \rho_d) > \epsilon | B\} = 0, \qquad \text{for all } \epsilon > 0.$$

Since $D(\rho, \rho_d) \leq 1$, for all $\rho \in \mathcal{S}$, we know that for all $\epsilon > 0$

$$\mathbb{E}[D(\rho_t, \rho_d)|B] \leq \mathbb{P}\{D(\rho_t, \rho_d) > \epsilon|B\} + \epsilon(1 - \mathbb{P}\{D(\rho_t, \rho_d) > \epsilon|B\}).$$

Hence,

$$\limsup_{t\to\infty} \mathbb{E}[D(\rho_t, \rho_d)|B] \leq \epsilon \qquad \text{for all } \epsilon > 0.$$

Thus, it is easy to see that

$$\lim_{t\to\infty} \mathbb{E}[D(\rho_t, \rho_d)|B] = 0.$$

Note that $\mathcal{L}D(\rho, \rho_d) = -\text{Tr}^2(i[H_b, \rho]\rho_d) \leq 0$ in $\mathcal{S}_{<1-\gamma/2}$, then by the Theorem 2.2 in [17], $D(\rho_t, \rho_d)$ converges a.s. for path remaining in $\mathcal{S}_{<1-\gamma/2}$. Therefore, by the dominated convergence theorem we have

$$\mathbb{E}[\lim_{t\to\infty} D(\rho_t, \rho_d)|B] = 0.$$

So

$$\mathbb{P}\{\lim_{t\to\infty} D(\rho_t, \rho_d) = 0|B\} = 1,$$

which completes the proof of this lemma. $\square$

## REFERENCES

[1] M. J. Biercuk, H. Uys, A. P. VanDevender, N. Shiga, W. M. Itano, J. J. Bollinger, Optimized dynamical decoupling in a model quantum memory, Nature, (2009), 458: 996-1000.

[2] J. F. Du, X. Rong, N. Zhao, Y. Wang, J. H. Yang, R. B. Liu, Preserving electron spin coherence in solids by optimal dynamical decoupling, Nature, (2009), 461: 1265-1268.

[3] A. D. O'Connell, M. Hofheinz, M. Ansmann, et al, Quantum ground state and single-phonon control of a mechanical resonator, Nature, (2010), 464: 697-703.

[4] G. Huang, T. J. Tarn, J. W. Clark, On the controllability of quantum-mechanical systems, J. Math. Phys., (1983), 24: 2608-2618.

[5] C. Altafini, Controllability properties for finite dimensional quantum Markovian master equations, J. Math. Phys., (2003), 44: 2357-2372.

[6] J. S. Li, N. Khaneja, Ensemble control of bloch equations, IEEE Trans. Autom. Control, (2009), 54: 528-536.

[7] R. Romano, D. D'Alessandro, Incoherent control and entanglement for twodimensional coupled systems, Phys. Rev. A, (2006), 73, 022323.

[8] R. Romano, D. D'Alessandro, Enviroment-mediated control of a quantum system, Phys. Rev. Lett., (2006), 97, 080402.

[9] N. Ganesan, T. J. Tarn, Quantum internal model principle: decoherence control, http://arxiv.org/abs/1004.0666

[10] R. S. Judson, H. Rabitz, Teaching lasers to control molecules, Phys. Rev. Lett., (1992), 68: 1500-1503.

[11] E. V. Wilson, Quantum control, Chemical & Engineering News, (2001), 79 (22): 38-39.

[12] W. S. Zhu, H. Rabitz, Closed loop learning control to suppress the effects of quantum decoherence, J. Chem. Phys., (2003), 118: 6751-6757.

[13] H. M. Wiseman, G. J. Milburn, Quantum Measurement and Control, Cambridge University Press, Cambridge, (2010).

[14] C. Ahn, A. C. Doherty, A. J. Landahl, Continuous quantum error correction via quantum feedback control, Phys. Rev. A, (2002), 65, 042301.

[15] H. M. Wiseman, A. C. Doherty, Optimal unravellings for feedback control in linear quantum systems, Phys. Rev. Lett., (2005), 94, 070405.

[16] R. van Handel, J. K. Stockton, H. Mabuchi, Feedback control of quantum state reduction, IEEE Trans. Autom. Control, (2005), 50: 768-780.

[17] M. Mirrahimi, R. van Handel, Stabilizing feedback control for quantum systems, SIAM J. Control Optim., (2007), 46: 445-467.

[18] M. R. James, H. I. Nurdin, I. R. Petersen, $H^\infty$ control of linear quantum systems, IEEE Trans. Automat. Control, (2008), 53: 1787-1803.

[19] H. I. Nurdin, M. R. James, I. R. Petersen, Coherent quantum LQG control, Automatica, (2009), 45: 1837-1846.

[20] A. Maalouf, I. R. Petersen, Coherent $H^\infty$ control for a class of linear complex quantum systems, Proc. American Control Conference, (2009), 1472-1479.

[21] F. Xue, S. X. Yu, C. P. Sun, Quantum control limited by quantum decoherence, Phys. Rev. A, (2006), 73, 013403.

[22] H. Mabuchi, Coherent-feedback quantum control with a dynamic compensator, Phys. Rev. A, (2008), 78, 032323.

[23] A. C. Doherty, K. Jacobs, G. Jungman, Information, disturbance and Hamiltonian quantum feedback control, Phys. Rev. A, (2001), 63, 062306.

[24] B. Qi, L. Guo, Is measurement-based feedback still better for quantum control systems? Systems & Control Letters, (2010), 59: 333-339.

[25] H. Pan, B. Qi, L. Guo, Comparisons between open-loop and feedback controls based on a coherent quantum control model, Proceedings of the 29th Chinese Control Conference, July, 29-31, 2010, Beijing, China.

[26] L. Guo, On critical stability of discrete-time adaptive nonlinear control, IEEE Trans. Autom. Control, (1997), 42: 1488-1499.

[27] L. L. Xie, L. Guo, How much uncertainty can be dealt with by feedback? IEEE Trans. Autom. Control, (2000), 45: 2203-2217.

[28] L. Guo, Exploring the capability and limits of the feedback mechanism, Proc. of the International Congress of Mathematicians, (2002), Vol. 3: 785-794 (invited lecture).

[29] R. van. Handel, J. K. Stockton, H. Mabuchi, Modeling and feedback control design for quantum state preparation, Journal of Optics B: Quantum and Semiclassical Optics, (2005), 7 (10): S179-S197.

[30] K. R. Parthasarathy, An Introduction to Quantum Stochastic Calculus, Birkhäuser, Berlin, (1992).

[31] L. Bouten, R. van. Handel, M. R. James. An introduction to quantum filtering, SIAM Journal on Control and Optimization, (2007), 46 (6): 2199-2241.

[32] R. S. Liptser, A. N. Shiryayev, Statistics of Random Processes I: General Theory, Springer-Verlag, (2001).

[33] V. P. Belavkin, Quantum stochastic calculus and quantum nonlinear filtering, J. Multivariate Anal., (1992), 42: 171-201.

[34] H. M. Wiseman, G. J. Milburn, Quantum theory of optical feedback via homodyne detection, Phys. Rev. Lett., (1993), 70, 548.

[35] S. G. Schirmer, X. Wang, Stabilizing open quantum systems by Markovian reservoir engineering, Phys. Rev. A, (2010), 81, 062306.

[36] H. M. Wiseman, S. Mancini, J. Wang, Bayesian feedback versus Markovian feedback in a two-level atom, Phys. Rev. A, (2002), 66, 013807.

[37] M. A. Nielsen, I. L. Chuang, Quantum Computation and Quantum Information, Cambridge University Press, Cambridge, (2000).

[38] I. Tsukamoto, Perturbation method for linear periodic systems III, Funkcialaj Ekvacioj, (1985), 28: 103-116.

[39] P. E. Protter, Stochastic Integration and Differential Equation, Springer-Verlag, Berlin, (2004).

[40] R. Merris, Graph Theory, Wiley-Inerscience, NewYork, (2001).

[41] Y. S. Chow, H. Teicher, Probability Theory: Independence, Interchangeability, Martingales, third ed., Springer-Verlag, New York, (1997).

[42] B. L. S. Prakasa Rao, Semimartingales and their Statistical Inference, Chapman & Hall/CRC, (1999).

[43] H. J. Kushner, Stochastic Stability and Control, Academic Press, NewYork, (1967).

[44] J. Combes, K. Jacobs, Rapid state reduction of quantum systems using feedback control, Phys. Rev. Lett., (2006), 96, 010504.